\documentclass[prb,twocolumn,showpacs,10pt]{revtex4}
\usepackage{graphicx,color}
\usepackage{dcolumn}
\usepackage{amsmath}
\usepackage{amsfonts}
\usepackage{fancyhdr}
\usepackage{epsfig}
\usepackage{epic}

\newcommand{\beq}{\begin{equation}}
\newcommand{\eeq}{\end{equation}}
\newcommand{\bqa}{\begin{eqnarray}}
\newcommand{\eqa}{\end{eqnarray}}
\newcommand{\nn}{\nonumber}

\newcommand{\Eq}[1]{Eq.~(\ref{#1})}

\newcommand{\dg}{^\dagger}

\newcommand{\smallfrac}[2]{\mbox{$\frac{#1}{#2}$}}
\newcommand{\half}{\smallfrac{1}{2}}
\newcommand{\bra}[1]{\langle{#1}|}
\newcommand{\ket}[1]{|{#1}\rangle}
\newcommand{\sch}{Schr\"odinger }

\newcommand{\ito}{It\=o\ }

\newcommand{\sq}[1]{\left[ {#1} \right]}
\newcommand{\cu}[1]{\left\{{#1} \right\}}
\newcommand{\ro}[1]{\left( {#1}\right)}
\newcommand{\an}[1]{\left\langle{#1}\right\rangle}
\newcommand{\st}[1]{\left|{#1}\right|}
\newcommand{\tr}[1]{\mathrm{Tr}\sq{ {#1}}}

    \newcommand{\D}[1]{{\cal D}\sq{{#1}}}
    \newcommand{\J}[1]{{\cal J}\sq{{#1}}}
    \newcommand{\A}[1]{{\cal A}\sq{{#1}}}
    
    \newcommand{\h}[1]{{\cal H}\sq{{#1}}}
    \newcommand{\s}{{\cal S}}
    \newcommand{\Ref}[1]{Ref.\ \onlinecite{#1}}
    \newcommand{\Fig}[1]{Fig.\ \ref{#1}}
    \newcommand{\Sec}[1]{Sec.\ \ref{#1}}
    \newcommand{\sig}{\hat{\sigma}}
    \newcommand{\E}[1]{{\rm E}\sq{#1}}
    
    \newcommand{\proj}[2]{\ket{#1}\bra{#2}}

\begin{document}

\title{Sensitivity and back-action in charge qubit measurements by a strongly 
coupled single-electron transistor}

\author{Neil P. Oxtoby}
\email{n.oxtoby@griffith.edu.au}

\author{H. M. Wiseman}
\email{h.wiseman@griffith.edu.au}

\author{He-Bi Sun}
\affiliation{%
Centre for Quantum Computer Technology,
Centre for Quantum Dynamics,
School of Science,
Griffith University,
Brisbane 4111,
Australia
}%

\begin{abstract}
We consider charge-qubit monitoring (continuous-in-time weak measurement) 
by a single-electron transistor (SET) operating in the sequential-tunneling 
r\'egime.  We show that commonly used master equations for this r\'egime are 
not of the Lindblad form that is necessary and sufficient for guaranteeing 
valid physical states.  
In this paper we derive a Lindblad-form master equation and a 
corresponding quantum trajectory model for continuous measurement of 
the charge qubit by a SET.  Our approach requires that the SET-qubit 
coupling be strong compared to the SET tunnelling rates.  
We present an analysis of the \textit{quality} of the qubit measurement 
in this model (sensitivity versus back-action).  
Typically, the strong coupling when the SET island is occupied causes 
back-action on the qubit beyond the quantum back-action necessary for its 
sensitivity, and hence the conditioned qubit state is mixed.  
However, in one strongly coupled, asymmetric r\'egime, 
the SET can approach the limit of an ideal detector with an almost pure 
conditioned state.  We also quantify the quality of the SET using more 
traditional concepts such as the measurement time and decoherence time, 
which we have generalized so as to treat the strongly responding r\'egime.

\end{abstract}

\date{\today} 

\pacs{73.23.Hk, 03.67.Lx}

\maketitle

\section{Introduction\label{sec:intro}}
The single-electron transistor\cite{AveLikJLTP86,FulDolPRL87} (SET) 
has been suggested\cite{SSPRB98,DevSchNAT00} as a device to measure the 
state of charge qubits, one proposal for the fundamental
elements of a quantum computer.\cite{QCQI}  
In solid-state systems it is well known that realistic measurement devices 
such as the SET and quantum point contact\cite{FieetalPRL93} (QPC) do not 
perform instantaneous measurements.  Rather, it is necessary to treat the 
measurement as a sequence of weak measurements, resulting in a continuous 
process in time.

During such a continuous measurement, the conditioned qubit state is 
gradually projected into an eigenstate in the measurement basis.  Here the 
conditioned state is one that an experimenter would calculate knowing the 
output of the detector. On average (that is, ignoring this information) the 
effect of the measurement is simply to decohere (that is, remove the 
coherences of) the qubit in the measurement basis. This quantum back-action 
is unavoidable because the coherences between the eigenstates (which indicate 
that the qubit is in a superposition of these eigenstates) must vanish for 
one or the other eigenstate to be realized.  For an ideal\cite{KorPRB99} or 
quantum-limited\cite{QMeas} detector, the decoherence is equal to the 
minimum back-action allowed by quantum mechanics given the information 
obtained from the detector.  Consequently, a qubit state conditioned 
upon the output of an ideal detector would remain pure if it began pure.  
This is the case for the usual model of the 
QPC.\cite{GurPRB97,KorPRB99,GoaMilWisSunPRB01}

In this paper we maintain the purity of the conditioned state as the 
ultimate quantifier of the quality of the measurement. However, it has 
been common to use a different quantifier, namely a comparison between 
the decoherence rate $\Gamma_{\rm d}$ and the measurement rate 
$\tau^{-1}_{\rm m}$. These are the characteristic rates of the 
decoherence and the measurement (gradual projection) processes 
discussed above.  (It is assumed that there is only one characteristic 
rate).  For so-called ``weakly responding''\cite{KorPRB99} detectors, 
the ratio of these rates is limited by\cite{DevSchNAT00}  
\begin{equation}
\label{eq:hup1}
\Gamma_{\rm d}\tau_{\rm m} \geq \frac{1}{2} ,
\end{equation}
with equality implying an ideal detector.  
While not the main focus of this paper, we show 
that in general the detector quality cannot be captured by such a 
simple ratio, which is sometimes referred to as the detector 
efficiency.\cite{PilButPRL02}  
For example, for a QPC it is easy to verify that 
$\Gamma_{\rm d}\tau_{\rm m}$ varies between 1 (in the 
strong-response limit) and 1/2 (in the weak-response limit).  This is 
discussed in Appendix \ref{appQPC}. 

To date, theoretical treatments%
\cite{SSPRB98,MakSchShnPRL00,KorPRB01b,MakSchShnRMP01,MakSchShn02} 
of charge-qubit continuous measurements by SETs operating in the 
sequential-tunneling mode have focussed on the weakly responding 
detector.  This is where the variation in the detector's output 
that depends on the qubit state is small compared to the average 
part.  We will quantify this later.  

In this paper we begin by showing that the master equations derived 
for the SET-monitored charge qubit in Refs. \onlinecite{SSPRB98}, 
\onlinecite{MakSchShnPRL00}, \onlinecite{MakSchShnRMP01}, 
\onlinecite{MakSchShn02}, and \onlinecite{GoaPRB04} are not of the 
Lindblad\cite{LinCMP76} form that is necessary and sufficient for 
guaranteeing valid physical states.  Specifically, we show that for 
short times the qubit state matrix (density matrix) may become 
non-positive.  Positivity of the state matrix is a fundamental 
requirement for a model to represent a real physical system.  
Consequently, their evolution (the qubit density matrix) can violate 
positivity and hence cannot possibly represent the state of a real 
physical system.  
As well as our Lindblad-form master equation, we also present the 
associated equation for the conditioned qubit state, known as a 
\textit{quantum trajectory equation}\cite{OpenSys,WisMilPRA93b,WisQSO96} 
or \textit{quantum filtering equation}.\cite{BelRAE80,BelQCM95}  
To derive our equations requires assuming strong SET-qubit coupling, 
as we will discuss.

Using our model we show that the strongly coupled SET can, in the 
r\'egime of strong response, approach operation at the quantum limit 
where the purity of the qubit state is preserved in a conditional 
measurement.  In this r\'egime the SET also approaches the quantum 
limit of efficiency for a charge qubit measurement given by \Eq{eq:hup1}. 

The paper is organized as follows.  The next section contains a 
brief discussion of the SET and charge qubit system.  
In \Sec{sec:previous} we discuss previous models for SET-monitored 
charge qubits and show that the master equation in this 
case\cite{SSPRB98} is not of the Lindblad form.  In \Sec{sec:quantify} 
we present new general definitions for the decoherence and 
measurement times that are valid for arbitrary coupling strength and 
response.  We present our quantum trajectory model that corresponds to 
charge qubit measurement by a strongly coupled SET in \Sec{sec:us}.  
In the same section we present quantitative analyses of our master 
equations (conditional and unconditional), focussing on comparing 
sensitivity and back-action in the charge qubit measurement.  
The paper is concluded in \Sec{sec:conclusion}.

\section{System\label{sec:system}}
In this section we outline a quite general model for a SET 
monitoring a charge qubit, independent of coupling strength at this stage.  
For specificity, we consider the double quantum dot (DQD) charge 
qubit,\cite{HayetalPRL03} but expect the results to hold for other 
realizations of a charge qubit.  A schematic of the DQD-SET system is shown 
in \Fig{fig:dqdset}.  We discuss this figure below.

For measurement purposes, sequential tunneling processes are the most 
important as they yield the largest SET currents.\cite{YaleSPIE03}  Thus, 
we assume that the dominant form of transport through the SET is sequential 
tunneling, with higher-order processes such as cotunneling exponentially 
suppressed.  In this r\'egime the so-called orthodox 
theory\cite{AveLikJLTP86,KorPRB94} applies.  This requires the SET tunnel 
junction resistances to be greater than the resistance quantum 
$h/e^2 \approx 26 \mathrm{k\Omega}$,\cite{vHoBeeStaSCT1992,DevSchNAT00} 
and the SET to be biased `near' a charge degeneracy point.\cite{YaleSPIE03}  
Higher-order processes become more important away from a charge degeneracy 
point, where sequential tunneling events become less energetically 
favorable.  There is a subtlety here for the strongly responding SET that 
we discuss when we introduce our model in \Sec{sec:us}.  

For low temperature $k_{\rm B}T$ and not-too-high bias voltage 
$eV = \mu_{\rm L}-\mu_{\rm R}$ (relative to the SET island charging energy 
$E_{\rm C}=e^2/2C$, where $C$ is the island's total capacitance), the SET 
island has only two possible charge configurations, $\aleph =0$ and 
$\aleph=1$.  
With these conditions satisfied, the gate voltage (which is modified by 
the qubit electron) controls the amount of current flow through the SET.  
This is what enables the SET to measure the qubit state.  
For convenience, we assume that electron transport occurs only in the 
source-drain direction due to the finite SET bias voltage $eV$ 
(see \Fig{fig:dqdset}).

\begin{figure}[ht]
\includegraphics[width=0.45\textwidth]{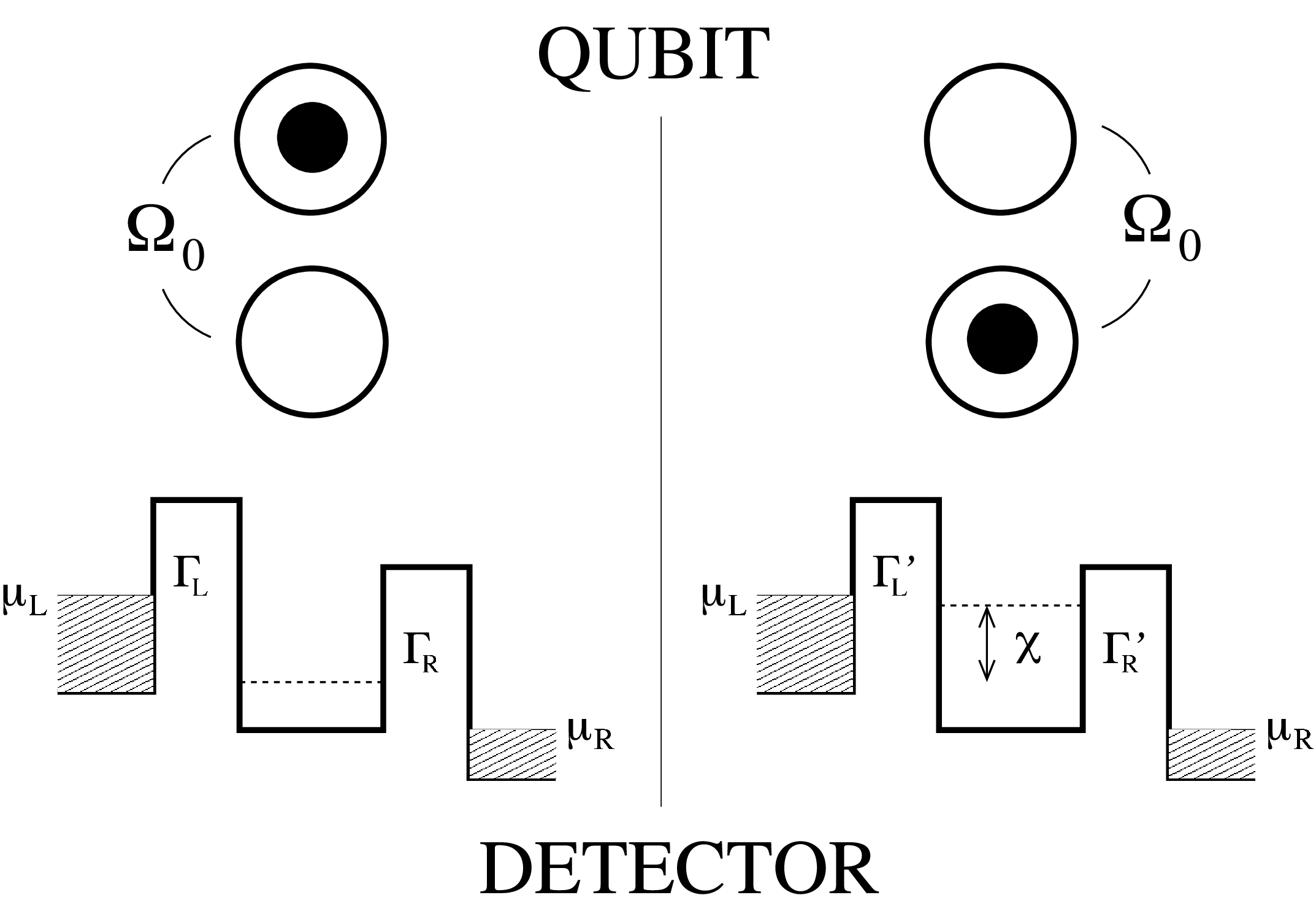}
\caption{\label{fig:dqdset}
Schematic of a double quantum dot (DQD) qubit and coupled SET.  A
single excess electron is shared by the DQDs.  When the near (far) QD
is occupied, the electron tunneling rates through the SET junctions
$j={\rm L,R}$ are denoted $\Gamma'_{j}$ ($\Gamma_{j}$).  
The charging energy of the SET island is increased by $\chi$ ($\hbar =1$) 
when the near QD is occupied.}
\end{figure}

The SET source and drain leads (reservoirs) are biased to Fermi levels 
$\mu_{\rm L}$ and $\mu_{\rm R}$, respectively.  
Associated with each of the qubit states is a tunneling process 
onto, and a tunneling process off, the SET island.  These processes 
occur at rates denoted $\Gamma_{\rm L}$, $\Gamma'_{\rm L}$ 
(source to island) and $\Gamma_{\rm R}$, $\Gamma'_{\rm R}$ 
(island to drain).  A dash indicates that the qubit electron is 
localized in the nearby QD, which we refer to as the 
``target''.\cite{WisetalPRB01}  
The average current through the SET is 
$I' = e\Gamma'_{\rm L}\Gamma'_{\rm R}/\ro{\Gamma'_{\rm L}+\Gamma'_{\rm R}}$ 
when the target dot is occupied, and 
$I = e\Gamma_{\rm L}\Gamma_{\rm R}/\ro{\Gamma_{\rm L}+\Gamma_{\rm R}}$ 
when the target dot is unoccupied.  
For later convenience, we now define the averages 
\mbox{%
$\bar{\Gamma}_{\rm L,R}\equiv \ro{\Gamma'_{\rm L,R}+\Gamma_{\rm L,R}}/2$} 
and differences 
\mbox{%
$\Delta\Gamma_{\rm L,R} \equiv \Gamma'_{\rm L,R}-\Gamma_{\rm L,R}$}.  

The Hamiltonian for a charge qubit with energy asymmetry $\varepsilon$ and 
tunnel-coupling strength $\Omega_0$ (see \Fig{fig:dqdset}) is 
$\hat{H}_{\rm qb} = \ro{\varepsilon \sig_{z} + \Omega_{0}\sig_{x}}/2$,
where $\sig_{x,z}$ are Pauli matrices.  We choose units such that $\hbar =1$.  
For $\varepsilon=0$, and in the absence of measurement, the qubit electron 
tunnels between the DQDs coherently (that is, existing in superpositions 
of the two localized states within the dots).  For characterizing the 
measurement quality in \Sec{sec:quantify} it is simpler (and 
commonplace\cite{SSPRB98,MakSchShnPRL00,MakSchShnRMP01,MakSchShn02}) 
to consider the limit of $\Omega_0 \rightarrow 0$, which we do.  
However, our master equations (both conditional and unconditional) are 
valid for non-zero $\Omega_{0}$.  When both the SET island and target 
dot are occupied, the system energy increases by the mutual charging 
energy, $\chi$.  This is described by the coupling Hamiltonian 
$\hat{H}_{\chi} = \chi\hat{\aleph}\otimes\hat{n}$, where 
$\hat{\aleph}=\hat{b}\dg\hat{b}$ and $\hat{n}\equiv (\sig_z+\hat{1})/2$ 
are the respective occupation number operators of the SET island and 
target dot.  For simplicity we do not show the standard Hamiltonians for 
the source and drain leads, since we trace the leads out of the total 
state matrix in order to consider only the SET and qubit in the 
Born-Markov approximation.  The Hamiltonian for the SET-DQD system (in 
the charge basis) is 
\begin{equation}
\hat{H} = \hat{1}\otimes \hat{H}_{\rm qb} + \hat{H}_{\chi}\ ,
\label{eq:Ham}
\end{equation}
where $\hat{1}$ is the $2\times 2$ identity matrix.  
It is simple to diagonalize the qubit Hamiltonian to find the 
computational/logical basis.\cite{SSPRB98}  However, we are interested 
in the measurement process which, for the purposes of this paper, is 
performed in the charge basis.

\section{Previous work\label{sec:previous}}
In this section we discuss previous work on the continuous monitoring of 
a charge qubit by a SET.  We demonstrate that the master equation of Ref. 
\onlinecite{SSPRB98} (and those of Refs. \onlinecite{MakSchShnPRL00} and 
\onlinecite{MakSchShnRMP01,MakSchShn02,GoaPRB04} since they are equivalent) 
can produce non-physical results and therefore is not of the Lindblad form.  
This master equation was derived independent of the coupling strength 
$\chi$, but the SET quality analysis always assumed weak coupling.  
First, we clarify the weak-response and weak-coupling assumptions.

Qualitatively, a weakly \textit{responding} detector is one in which 
the variation in the detector's output that depends 
on the qubit state is small compared to the average part.  
Korotkov\cite{KorPRB99,KorPRB01b,Kor03} expresses 
this quantitatively in terms of the current output as
\begin{equation}
\st{\Delta I} \ll \bar{I} ,
\label{eq:weakresponse}
\end{equation}
where $\Delta I \equiv I-I'$ and $\bar{I} \equiv \ro{I'+I}/2$.  
Note that $\st{\Delta I} \leq 2\bar{I}$, so that strong response is 
$\st{\Delta I} \approx 2\bar{I}$.

The strength of the SET-qubit \textit{coupling} is quantified by $\chi$ 
(see \Sec{sec:system}).  When analyzing the quality of the SET, the 
authors of Refs. \onlinecite{SSPRB98}, \onlinecite{MakSchShnPRL00}, 
\onlinecite{MakSchShnRMP01} and \onlinecite{MakSchShn02} make the 
weak coupling assumption 
\begin{equation}
\label{eq:weakcoupling}
\chi \ll \bar{\Gamma}_{\rm L} + \bar{\Gamma}_{\rm R} .
\end{equation}
In Ref. \onlinecite{MakSchShnRMP01} this assumption was made ``for 
definiteness''.  
It leads to a major simplification in that the SET island charge fluctuation 
spectrum is `white' over all frequencies of interest.  The variations in the 
SET tunneling rates are proportional to 
$\chi$,\cite{SSPRB98,MakSchShnPRL00,KorPRB01b,MakSchShnRMP01,MakSchShn02} 
so that weak coupling implies weak response.  However, a strongly coupled 
detector can have strong or weak response.

\subsection{Unconditional master equation\label{sec:SSPRB98}}
The unconditional (or nonselective, or ensemble average) master equation 
of \Ref{SSPRB98} is obtained by averaging over, or ignoring, the 
measurement record (the number of electrons that have tunneled into the 
SET drain).  This is achieved by setting $k=0$ in equations (20)--(27) of 
that paper.  To perform a quantitative analysis of these equations, it is 
imagined that the Josephson coupling energy can be ``turned off'' during 
the measurement (the equivalent Hamiltonian parameter from \Sec{sec:system} 
is $\Omega_0$).  Using our notations for clarity, the resulting 
unconditional master equation [equations (31) and (35) of that paper] 
can be written as
\begin{subequations}
\begin{eqnarray}
\label{eq:0_1}
\dot{\rho}^0_{11} & = & - \Gamma'_{\rm L}\rho^0_{11} 
+ \Gamma'_{\rm R}
\rho^1_{11}\ , \\
\label{eq:0_2}
\dot{\rho}^0_{22} & = & - \Gamma_{\rm L}\rho^0_{22}
+ \Gamma_{\rm R}\rho^1_{22}\ , \\
\label{eq:0_3}
\dot{\rho}^0_{12} & = & i\varepsilon \rho^0_{12}
- \bar{\Gamma}_{\rm L}\rho^0_{12} + \bar{\Gamma}_{\rm R}\rho^1_{12}\ , 
\end{eqnarray}
\label{eqs:SSme0}
\end{subequations}\begin{subequations}
\begin{eqnarray}
\label{eq:1_1}
\dot{\rho}^1_{11} & = & \Gamma'_{\rm L}\rho^0_{11} 
- \Gamma'_{\rm R}\rho^1_{11}\ , \\
\label{eq:1_2}
\dot{\rho}^1_{22} & = & \Gamma_{\rm L}\rho^0_{22}
- \Gamma_{\rm R}\rho^1_{22}\ , \\
\label{eq:1_3}
\dot{\rho}^1_{12} & = & i\ro{\varepsilon + \chi} \rho^1_{12}
- \bar{\Gamma}_{\rm R}\rho^1_{12}
+ \bar{\Gamma}_{\rm L}\rho^0_{12} 
\ .
\end{eqnarray}\label{eqs:SSme1}
\end{subequations}%
Here $\rho_{\rm i,j}^{\aleph}$ is the element in the $i$th row and $j$th 
column of the qubit state matrix conditioned by the SET state 
$\aleph = 0,1$ (the occupation number of the relevant SET island level).  
The states are normalized so that $P^{\aleph} \equiv \tr{\rho^{\aleph}}$ 
is the probability that the SET state is $\aleph$.  Thus the average 
qubit state is $\rho^{0} + \rho^{1}$.

We now demonstrate that Eqs. (\ref{eqs:SSme0}) and 
(\ref{eqs:SSme1}) can produce a non-positive, and therefore non-physical, 
qubit state matrix.  We begin by noting that the following inequality is 
satisfied by all positive $2\times 2$ matrices.
\begin{equation}
\label{eq:ineq}
\rho^{\phantom{0}}_{11}\rho^{\phantom{0}}_{22} \geq
\st{\rho^{\phantom{0}}_{12}}^2\ .
\end{equation}
Now consider the short-time solution for $\rho^0$ [Eqs. (\ref{eqs:SSme0})] 
when the SET is initially occupied ($\rho^0 = 0$ at $t=0$) and the qubit is 
in the following superposition state: 
$\rho^1_{11} = \rho^1_{22} = \rho^1_{12} =\frac{1}{2}$.  
After an infinitesimally short time $\delta t$, we have
\begin{eqnarray}
\rho^0_{11} & \approx & \half\Gamma'_{\rm R} \delta t \ ,\nn \\
\label{eq:solrho0_2}
\rho^0_{22} & \approx & \half\Gamma_{\rm R} \delta t \ ,\\
\rho^0_{12} & \approx & \half\bar{\Gamma}_{\rm R} \delta t\ .\nn
\end{eqnarray}
Substituting these into \Eq{eq:ineq}, canceling factors of
$\ro{\delta t}^2/4$ on both sides, and rearranging gives
\begin{equation}
\st{\Delta \Gamma_{\rm R}}^{2} \leq 0\ ,
\label{eq:false}
\end{equation}
which is false for all values of $\Gamma'_{\rm R}\neq \Gamma_{\rm R}$. 
The case of $\Gamma'_{\rm R} = \Gamma_{\rm R}$ corresponds to there 
being \textit{no response} from the SET for a change in the qubit state, 
i.e. no measurement is being performed.\footnote{See also the paragraph 
following Eq. (30) of \Ref{SSPRB98}.}  
The result (\ref{eq:false}) shows that the inequality (\ref{eq:ineq}) is 
not satisfied, positivity is not preserved, and the master equation of 
\Ref{SSPRB98} is not of the Lindblad form.  This is also true of the 
equivalent master equations presented in Refs. \onlinecite{MakSchShnPRL00}, 
\onlinecite{MakSchShnRMP01}, \onlinecite{MakSchShn02} and 
\onlinecite{GoaPRB04}.   
Numerical results show that the violation of \Eq{eq:ineq} occurs only 
for short times (compared to the 
characteristic qubit decoherence and measurement times).  However, since 
the qubit state matrix for later times is determined by the (non-positive) 
state matrix for short times, one cannot be confident in the ability of 
the state matrix at later times to describe a real physical system.  
Therefore, we maintain that the master equations of Refs. 
\onlinecite{SSPRB98}, \onlinecite{MakSchShnPRL00}, 
\onlinecite{MakSchShnRMP01}, \onlinecite{MakSchShn02} and 
\onlinecite{GoaPRB04} cannot be relied upon at any time.

\section{Quantifying the Sensitivity and Back-action\label{sec:quantify}}
In the previous section we showed that previous master equations for the 
SET-monitored charge qubit are not of the Lindblad form.  
In \Sec{sec:us} we derive a Lindblad-form master equation (and 
``unravel''\cite{OpenSys} it into a conditional master equation), which 
requires strong coupling between the SET and the qubit.  
In the strong coupling r\'egime it is natural for the detector also 
to be strongly responding since, as mentioned, the coupling energy 
largely determines the detector's response.  
The more traditional measures of sensitivity and back-action in this 
type of measurement 
are\cite{SSPRB98,MakSchShnPRL00,KorPRB01b,MakSchShnRMP01,MakSchShn02} 
the measurement time $\tau_{\rm m}$ and decoherence rate $\Gamma_{\rm d}$, 
respectively.  Previous definitions of these quantities assumed weak 
detector response, either explicitly or implicitly.  Therefore it is 
necessary to use more general definitions, which we give in this section.

\subsection{Decoherence Time\label{sec:deco}}
Previously the decoherence rate $\Gamma_{\rm d}$ was defined as the rate 
at which the coherences in the qubit decay according to the unconditional 
master equation.  For charge qubit measurements by a QPC, this 
description is unambiguous: there is one decay rate for the coherences 
which can be obtained\cite{GurPRB97} by inspection of the master equation.  
This is not the case for the SET because there are two qubit state 
matrices, namely $\rho^0$ and $\rho^1$, conditioned by the SET island 
state.  Our master equation (\ref{eq:ME}) gives a rate of decay for 
each of these conditional qubit coherences, neither of which 
\textit{alone} accurately represents the average decoherence of 
the qubit state $\rho = \rho^0 + \rho^1$.  
In Refs. \onlinecite{SSPRB98}, \onlinecite{MakSchShnPRL00}, 
\onlinecite{MakSchShnRMP01}, and \onlinecite{MakSchShn02}, the qubit 
decoherence rate is defined as the slowest of these two rates 
under the premise that the qubit state cannot collapse any faster.  
This therefore represents an \textit{over-estimate} of the quality 
of the measurement device, as it minimizes \Eq{eq:hup1}.

In an effort to more precisely quantify the decoherence, we find a 
characteristic qubit decoherence time by borrowing a technique from 
the field of quantum optics.  In a typical quantum optics scenario, 
where the detector signal consists of absorption of photons emitted 
by a quantum system, the coherence of the source is defined by the 
Glauber coherence function.\cite{GlaPR63a}  For a two-level quantum 
system (such as the two-level atom) this is given in terms of the 
steady-state, two-time correlation function 
$g(\tau) = \an{\sig\dg(\tau)\sig0}_{\rm ss}$.  Here 
$\sig = \ket{0}\bra{1}$ is the lowering operator for the atom (in 
the Heisenberg picture).  The expression involves the lowering 
operator because the detection involves absorption.  

In the solid-state context where the detection does not 
``lower'' the DQD state it is more appropriate to consider an 
expression that treats $\ket{0}$ and $\ket{1}$ symmetrically.  
The obvious expression is
\begin{equation}
g(\tau) \equiv \an{\sig\dg(\tau)\sig 0}_{\rm ss} +
\an{\sig(\tau)\sig\dg0}_{\rm ss}^{*}\ .
\label{eq:gdqd}
\end{equation} 
In the \sch picture, this coherence function can be expressed as 
\begin{eqnarray}
g(\tau) &=&
\tr{\sig\dg{\rho}^{-}(\tau)} + \tr{\sig{\rho}^{+}(\tau)}^{*}
\nn\\
&\equiv& {\rho}^{-}_{21}(\tau) + \sq{{\rho}^{+}_{12}(\tau)}^*
\ ,
\label{eq:gSP}
\end{eqnarray}
where 
${\rho}(\tau) = \exp \ro{\mathcal{L}_{\rho}\tau}{\rho}(0)$, 
${\rho}^{+}(0)\equiv \sig\dg\rho_{\rm ss}$, and 
${\rho}^{-}(0)\equiv \sig\rho_{\rm ss}$.  The Liouvillian 
$\mathcal{L}_{\rho}$ is the qubit time evolution superoperator 
[analogous to $\mathcal{L}$ in \Eq{eq:ME}].

The coherence function thus defined has the property that 
it is normalized in the sense that 
$g(0) \equiv 1$ and $\lim_{\tau \to \infty} g(\tau) = 0$.  
Therefore we can now define the decoherence time (which means 
the same as the coherence time) as\cite{QNoise}
\begin{equation}
\tau_{\rm d} \equiv \int_{0}^{\infty} g(\tau) d\tau .
\label{eq:decotimedef}
\end{equation}
The decoherence rate is simply $\Gamma_{\rm d} = \tau_{\rm d}^{-1}$.  
For an exponentially decaying coherence it is easy to verify that 
$\Gamma_{\rm d}$ is the exponential decay coefficient.
For example, for the QPC master equation of \Ref{GoaMilWisSunPRB01}, this coherence 
function technique yields the same decoherence rate as the QPC models of Refs. 
\onlinecite{KorPRB99}, \onlinecite{GurPRB97} and \onlinecite{GoaMilWisSunPRB01} 
(see Appendix \ref{appQPC}).

\subsection{Measurement Time\label{sec:meas}}
Previously\cite{SSPRB98,MakSchShnPRL00,MakSchShnRMP01,MakSchShn02} 
the measurement time $\tau_{\rm m}$ was defined as follows.  
Keeping track of the number of electrons $N_{\rm R}(t)$ that have 
tunneled into the drain during the measurement process (while ignoring 
the information contained by electrons tunneling from the source onto 
the island) allows tracking of the dynamics of the probability 
distribution for this number, $P(N_{\rm R},t)$.  Gaussian statistics were 
assumed for $N_{\rm R}$, thereby implicitly assuming weak detector response 
(many SET tunneling events occur before the qubit states are 
distinguishable).  After starting the measurement as a delta function at 
$N_{\rm R}=0$, $P(N_{\rm R},t)$ displays two peaks (corresponding to the 
two qubit states) which drift linearly in time to positive values of 
$N_{\rm R}$ with velocities $\Gamma \equiv I/e$ and $\Gamma' \equiv I'/e$, 
respectively.  The widths of the peaks broaden due to counting statistics, 
growing as $\sqrt{f\Gamma t}$ and $\sqrt{f'\Gamma't}$, 
respectively.\cite{MakSchShn02}  In Refs. \onlinecite{MakSchShnPRL00} and 
\onlinecite{MakSchShnRMP01} these are given with an extra factor of 
$\sqrt{2}$ as $\sqrt{2f\Gamma t}$ and $\sqrt{2f'\Gamma' t}$.  
Here the Fano factors are 
$f = \ro{{\Gamma}_{\rm L}^2 + {\Gamma}_{\rm R}^2}
/\ro{{\Gamma}_{\rm L}+{\Gamma}_{\rm R}}^2$ 
and 
$f' = \ro{{\Gamma}_{\rm L}^{\prime2} + {\Gamma}_{\rm R}^{\prime2}}
/\ro{{\Gamma}'_{\rm L}+{\Gamma}'_{\rm R}}^2$.  
The peaks emerge from the broadened distribution when the separation of 
the distributions is larger than their widths,  
$\st{\Gamma - \Gamma'}t \geq \ro{\sqrt{f\Gamma t} + \sqrt{f'\Gamma't}}$.  

Fundamentally, the measurement time relates to distinguishing the qubit 
states using information gained from the detector.  This obviously 
involves the conditioned state of the qubit.  Since the quantum 
trajectory equation is by definition the optimal way to process the 
detector current to get information about the qubit, it should be used 
to define the measurement time.  Our quantum trajectory equation is 
given in \Sec{sec:us}.

In the measurement basis we can, quite generally, define an 
uncertainty function for the qubit state as the conditional variance 
in $\sig_z$:
\begin{subequations}
\begin{eqnarray}
V(t) &=& \E{ \an{\sig_z^2}_{\rm c}(t) - \an{\sig_z}_{\rm c}^2(t)} \\
\label{eq:V1}
&=& 1 - \E{z_{\rm c}^2(t)} .
\label{eq:V2}
\end{eqnarray}
\label{eqs:V}
\end{subequations}
Here the subscript c denotes conditional variables, 
\mbox{E[ $\cdot$ ]} denotes a classical expectation value, and 
$\an{ \cdot }$ denotes a quantum average (we use $z$ as shorthand 
for $\an{\sig_z}$).  The initial state is the unconditional steady 
state, that is, the long-time limit of the master equation given 
in \Sec{sec:us}.

Provided that the qubit Hamiltonian commutes with the measured 
observable, i.e. $\sig_{z}$, the measurement will not disturb the 
qubit populations (but will necessarily affect the qubit coherence).  
Thus, for a non-disturbing measurement we require $\Omega_{0}=0$.  
In this case, $V(t)$ will have the properties that $V(0) = 1$, and 
$\lim_{\tau \to \infty} V(\tau) = 0$.  That is, we start off with no 
knowledge of $\sig_z$ and end up with complete knowledge.  Thus, in 
analogy with \Eq{eq:decotimedef}, it can be used to give a general 
definition of the measurement time as 
\begin{equation}
\label{eq:meastimedef}
\tau_{\rm m} \equiv \int_{0}^{\infty} V(\tau) d\tau .
\end{equation}
However, because $V(t)$ arises from stochastic evolution, it is not 
generally possible to obtain an analytical expression for $V(\tau)$, 
unlike $g(\tau)$, which arises from the deterministic evolution of the
master equation.  If $V(t)$ decays exponentially with time, then 
$\tau_{\rm m}$ can be evaluated as 
\begin{equation}
\label{eq:dotV0}
\tau_{\rm m}^{-1} = - \left.\frac{d V(t)}{dt}\right|_{t=0} .
\end{equation}
This equation (expressed differently) was first used in 
Ref.~\onlinecite{WisetalPRB01}.  It has the advantage that it can 
be evaluated analytically.  (In Appendix \ref{appQPC} we report this 
calculation for the QPC.)  As we will see later, for the SET in a 
wide range of parameter r\'egimes, the exponential approximation 
for $V(\tau)$ is a good one, so that \Eq{eq:dotV0} can be used.

\section{Strongly coupled SET\label{sec:us}}
Using the quantum trajectory approach developed in quantum 
optics,\cite{OpenSys} we describe both the 
conditional\cite{KorPRB99,WisetalPRB01} and better-known 
unconditional\cite{GurPRB97} time evolution of the charge qubit state 
whilst undergoing continuous measurement by a SET.  
In order to consider the tunneling processes through the SET as 
distinguishable, that is, in order to ignore the finite widths 
$\Gamma_{\rm L,R}$ and $\Gamma'_{\rm L,R}$ of these tunneling 
processes, requires\cite{BeePRB91}
\begin{equation}
\chi \gg \bar{\Gamma}_{\rm L} + \bar{\Gamma}_{\rm R}\ .
\label{eq:strongcoupling}
\end{equation}
This assumption of a strongly coupled SET is critical to our ability to 
derive a master equation using our approach.  
As mentioned in \Sec{sec:system}, the qubit energy asymmetry is shifted 
by $\chi$ when both the SET island and the target dot are occupied.  For 
the strongly coupled SET, this causes a large shift ($\propto \chi$) 
in the qubit Rabi frequency ($\Omega = \sqrt{\Omega_0^2 + \varepsilon^2}$).  
One could conceivably posit that this situation cannot be considered as 
measurement as it has such a large effect on the qubit state.  However, 
because this energy shift is in the same basis as the measurement, it 
affects only the qubit coherence and not the populations.  
We analyze this below.

\begin{figure}[ht]
\includegraphics[width=0.5\textwidth]{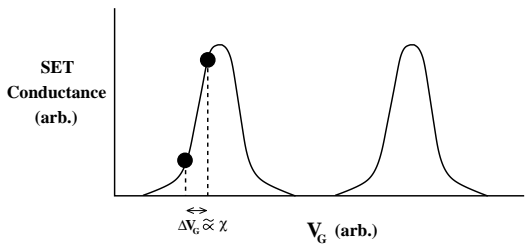}
\caption{Hypothetical SET conductance vs SET gate voltage.  
The two dots show the SET conductance corresponding to the two 
qubit eigenstates for the case of strong response.
\label{fig:response}}
\end{figure}

Note that the time to determine the qubit state (the measurement time) 
is related to the time it takes to distinguish between the two 
corresponding currents through the SET.  Intuitively, larger separation 
between these two currents decreases the measurement time.  Indeed, this 
leads to the prediction that the strongly \textit{responding} SET would 
enable faster charge qubit measurement than its weakly responding 
counterpart.  Qualitatively, the largest separation of these currents 
occurs when tunneling through the SET switches between on and off for 
the two qubit states.  When the SET is operated near this so-called 
``switching point'', thermal fluctuations and higher-order processes 
such as cotunneling become the leading contribution to the 
current.\cite{SSPRB98}  As these processes yield very small 
currents,\cite{YaleSPIE03} they may well be negligible in a laboratory 
setting.  Nevertheless, we assume that the SET conducts for both qubit 
states to ensure that higher-order processes are negligible.  Figure 
\ref{fig:response} shows this assumption schematically on a hypothetical 
SET conductance-oscillation plot.  The qubit state shifts the gate 
voltage by an amount that changes the SET conductance significantly for 
the case of strong response, while remaining always in a conducting state.

We represent the combined state of the SET island and qubit by the 
composite state matrix
\begin{equation}
G(t) = \proj{0}{0}_{\rm SET}\otimes \rho^{0}(t) + 
 \proj{1}{1}_{\rm SET}\otimes \rho^{1}(t) \ ,
\label{eq:G}
\end{equation}
where $\rho(t)=\tr{G(t)}=\rho^{0}(t)+\rho^{1}(t)$ is the qubit state 
matrix, the superscript again referring to the SET island occupation 
($\aleph=0$ or $1$).  
We reiterate that, although we consider the state of the SET island 
in a composite density matrix with the qubit, no charge superposition
states exist on the SET island.  We present master equations for $G(t)$ 
for compactness, but could equally well present coupled master equations 
for the qubit states $\rho^0(t)$ and $\rho^1(t)$ as in Refs. 
\onlinecite{MakSchShnPRL00} and 
\onlinecite{MakSchShnRMP01,MakSchShn02,GoaPRB04}, for example.

\subsection{Unconditional master equation\label{sec:average}}
Here we present our unconditional master equation for 
the state matrix $G(t)$ of the system (qubit plus SET island).  We 
will first outline the details of its derivation, which closely 
follows the derivation in the appendix of \Ref{WisetalPRB01}.  
We first consider the state matrix $W(t)$ for the combined system 
(qubit, SET and leads).  The leads are treated as perfect Fermi 
thermal reservoirs with very fast relaxation constants.  Each lead 
then remains in thermal equilibrium at its respective chemical 
potential.  We define a time interval $\Delta t$ that is long 
compared to the relaxation time of the leads, but short 
(infinitesimal) compared to the dynamics of the SET island and qubit.  
Transforming to an interaction picture with respect to $\hat{H}$ 
[\Eq{eq:Ham}] leaves only tunnel-coupling terms (between the leads 
and the SET island) in the interaction picture Hamiltonian 
$\hat{H}_{\rm i}$.  
The change in $W(t)$ from time $t$ to $t+\Delta t$ to second 
order in the tunnel-coupling energy is then given by
\begin{eqnarray}
W(t+\Delta t) &=& W(t) -i\Delta t \sq{\hat{H}_{\rm i},W(t)} \nn \\
&& -\Delta t \int_{t}^{t+\Delta t} dt' 
\sq{\hat{H}_{\rm i}(t'),\sq{\hat{H}_{\rm i}(t'),W(t')}} .\nn\\
\label{eq:secondorder}
\end{eqnarray}
We now make a Markov approximation for the instantaneous relaxation 
of the source (L) and drain (R) leads so that 
$W(t) = G(t)\otimes \rho_{\rm L} \otimes \rho_{\rm R}$, and we find 
the resulting evolution equation for $G(t)$ by tracing over the leads.  
Assuming that all tunneling occurs from source to drain 
($k_{\rm B}T \ll eV$) and moving back to the \sch picture, we obtain 
the unconditional master equation for $G(t)$ as
\begin{eqnarray}
\dot{G} &=& -i \sq{\hat{H},G}
+ \Gamma'_{\rm L}\D{\hat{b}\dg\otimes \hat{n}}G
+ \Gamma'_{\rm R}\D{\hat{b}\otimes \hat{n}}G \nn \\
&&+ \Gamma_{\rm L}\D{\hat{b}\dg\otimes (\hat{1}-\hat{n})}G
+ \Gamma_{\rm R}\D{\hat{b}\otimes (\hat{1}-\hat{n})}G \nn \\
&\equiv& {\cal L} G , \label{eq:ME}
\end{eqnarray}
which is explicitly of the Lindblad form, and we have omitted time 
arguments of $G$ for brevity.  We remind the reader that 
$\hat{b} = \ket{0}\bra{1}_{\rm SET}$, 
$\hat{n}=\ro{\sig_{z}+1}/2$ is the target dot occupation operator, 
and $\hat{H}$ is given by \Eq{eq:Ham}.  
In deriving \Eq{eq:ME}, our assumption of strong SET-qubit coupling 
($\chi \gg \bar{\Gamma}_{\rm L} + \bar{\Gamma}_{\rm R}$) is the main 
difference between our assumptions and those of the previous models.%
\cite{SSPRB98,MakSchShnPRL00,MakSchShnRMP01,MakSchShn02,GoaPRB04}  
This assumption is essential to our ability to derive 
our master equation, as it allows each lead to be treated as two 
independent baths that differ in energy by $\hbar \chi$.  
The Lindblad superoperator ${\cal D}$ in \Eq{eq:ME} represents the 
dissipative, irreversible part of the qubit evolution --- the 
measurement-induced decoherence.  It is defined as\cite{WisMilPRA93b} 
\begin{equation}
\label{eq:D}
\D{\hat{x}}G \equiv \J{\hat{x}}G - \A{\hat{x}}G\ ,
\end{equation}
where $\J{\hat{x}}G \equiv \hat{x} G\hat{x}\dg$, and 
$\A{\hat{x}}G\equiv\half\ro{\hat{x}\dg\hat{x}G+G\hat{x}\dg\hat{x}}$.  
Goan\cite{GoaPRB04} expressed the master equation of Eqs. 
(\ref{eqs:SSme0})--(\ref{eqs:SSme1}) in a similar form to \Eq{eq:ME}.  
The crucial differences, i.e. the non-Lindblad terms in Eqs. 
(\ref{eqs:SSme0})--(\ref{eqs:SSme1}), are shown in Appendix 
\ref{sec:goanme}.

\begin{figure}[ht]
\includegraphics[width=0.45\textwidth]{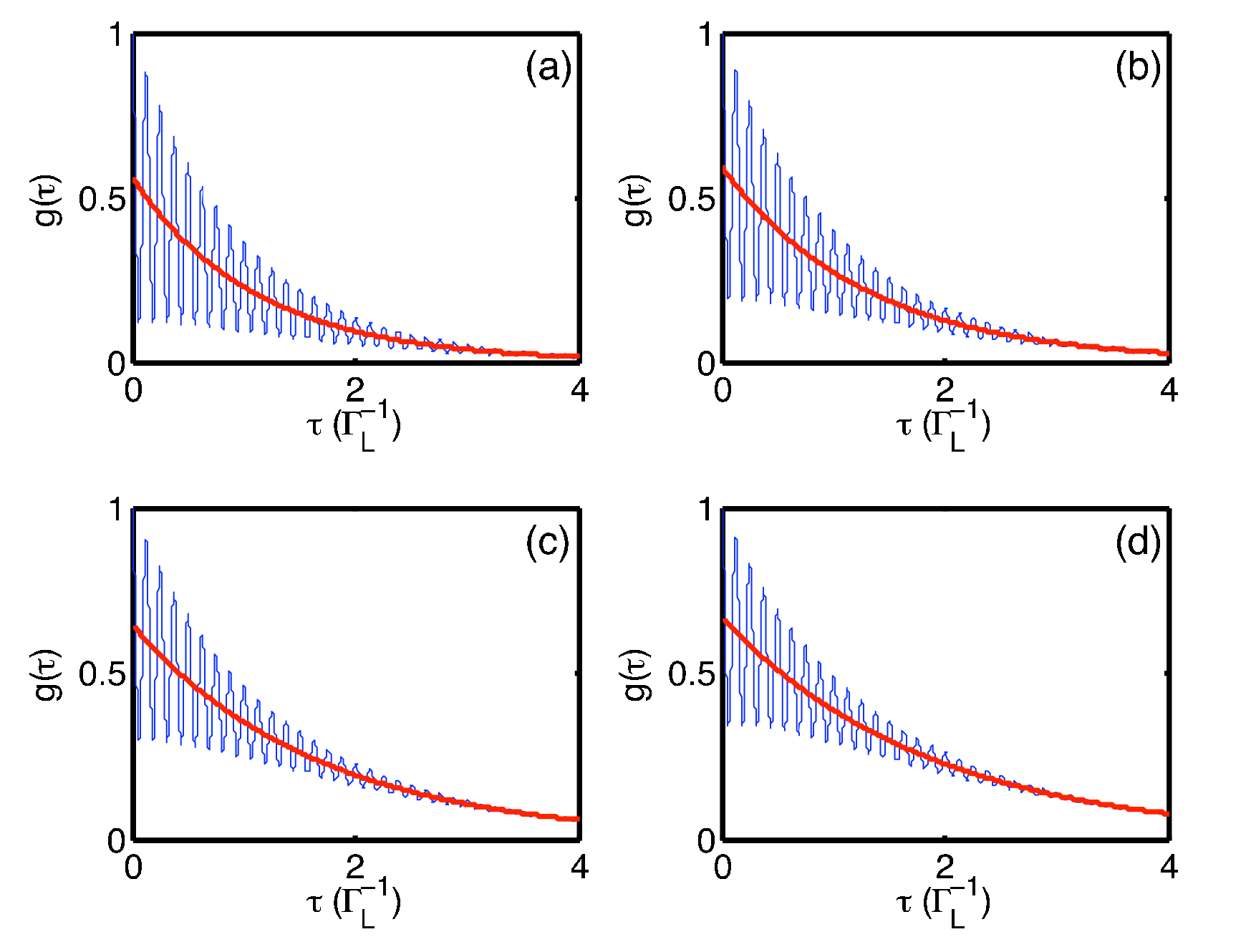}
\caption{Decoherence of a charge qubit undergoing continuous measurement 
by a strongly responding SET.  The rapid oscillations are too fast to be 
observable --- the thick line shows the observable result.  
From (a) to (d): 
$\st{\Delta I}/\bar{I} = 1/10$, $1/3$, $1$, and $3/2$.  Other parameters were $\Omega_0=0.01\Gamma_{\rm L}$, $\chi=50\Gamma_{\rm L}$, 
$\Gamma_{\rm R}=1.1\Gamma_{\rm L}$, $\Gamma'_{\rm R}=1.2\Gamma_{\rm L}$. 
\label{fig:g}}
\end{figure}

The unconditional master equation (\ref{eq:ME}) describes the system 
evolution when the measurement result is ignored or averaged over.  
When quantifying the SET measurement quality, we assume a non-disturbing 
measurement as noted in the paragraph preceding \Eq{eq:meastimedef}.  
That is, we assume $\Omega_{0}=0$ (as was assumed in previous 
quantitative analyses\cite{SSPRB98,MakSchShnPRL00,MakSchShnRMP01,MakSchShn02} 
of the SET measurement quality).  
In order for our master equation to have a unique steady-state solution, 
we require nonzero $\Omega_{0}$.  Thus, for numerical simulations we use 
$\Omega_0 = 0.01\Gamma_{\rm L}$.  Note however, that \Eq{eq:ME} is valid 
for any $\Omega_{0}\ll\chi$.  
The numerical solution of the master equation (\ref{eq:ME}) gives oscillating, 
decaying off-diagonal elements of the qubit state matrix.  We quantify this 
by the coherence function $g(\tau)$ given in \Eq{eq:gSP}.  For SET monitoring 
of the qubit, ${\cal L}$ acts on the qubit-plus-SET state $G$, so $\rho$ must 
be replaced by $G$ in the coherence function of \Eq{eq:gSP}.  The resulting 
$g(\tau)$ is shown in \Fig{fig:g}.  

The four subplots in \Fig{fig:g} correspond to increasing values of SET 
response $\st{\Delta I}/\bar{I}$, which is bounded above by $2$.  The 
strength of the SET response was modified by varying $\Gamma'_{\rm L}$ --- 
the slowest rate of tunneling onto the SET island.  
The initial qubit state is completely mixed:
\begin{equation}
\rho_{\rm ss} = \frac{1}{2}\hat{1} ,
\label{eq:qubitsteadystate}
\end{equation}
and the SET island is initially in the corresponding steady-state, which is 
represented by the island occupation probability $P^{1} = \tr{\rho^{1}}$ 
(with $P^{0} = 1-P^{1}$) as
\begin{equation}
P^{1}_{\rm ss} = \frac{\bar{\Gamma}_{\rm L}}
{\bar{\Gamma}_{\rm L}+\bar{\Gamma}_{\rm R}} .
\label{eq:SETsteadystate}
\end{equation}
The strongest SET response [in Fig. \ref{fig:g} (d)] corresponds to 
\begin{equation}
0 < \Gamma'_{\rm L} \ll \Gamma_{\rm L} \sim \Gamma_{\rm R} \sim
\Gamma'_{\rm R} ,
\label{eq:regime}
\end{equation}
where we assume that $\Gamma_{\rm L}'$ is large enough in order to ignore 
higher-order processes such as cotunneling.

The rapid oscillations in \Fig{fig:g} are due to the strong SET-qubit 
coupling given in \Eq{eq:strongcoupling}.  Thus, these oscillations are 
very fast compared to the tunneling rates.  As a consequence, even if an 
individual tunneling event could be detected, it would be impossible to 
know when it occurred with sufficient accuracy to allow the phase of the 
oscillation to be known.  This is a direct result of the strong coupling 
assumption (\ref{eq:strongcoupling}) required for our Lindblad-form 
master equation.  In the observable qubit state, the rapid oscillations 
are effectively averaged over. This produces the observable decoherence 
given by the bold line in \Fig{fig:g}.  

The observable state can be obtained from $G(t)$ by moving to an 
interaction picture with respect to $\hat{H}_{\chi}$, unique from the 
interaction picture mentioned in the derivation of \Eq{eq:ME}, and making 
a rotating wave approximation by dropping all rapidly oscillating terms 
(those involving $\chi$).  The result is the following master equation 
\begin{eqnarray}
\dot{\widetilde{G}} &=& -i \sq{\widetilde{H}_{\rm I},\widetilde{G}}
+ \Gamma'_{\rm L}\D{\hat{b}\dg\otimes \hat{n}}\widetilde{G}
+ \Gamma'_{\rm R}\D{\hat{b}\otimes \hat{n}}\widetilde{G} \nn \\
&&+ \Gamma_{\rm L}\D{\hat{b}\dg\otimes (\hat{1}-\hat{n})}\widetilde{G}
+ \Gamma_{\rm R}\D{\hat{b}\otimes (\hat{1}-\hat{n})}\widetilde{G} \nn \\
&\equiv& \widetilde{\mathcal{L}} \widetilde{G} , 
\label{eq:effME}
\end{eqnarray}
where the observable state $\widetilde{G}$ 
is given by
\begin{equation}
\widetilde{G} = 
  \proj{0}{0}_{\rm SET}\otimes \tilde{\rho}^{0} 
+ \proj{1}{1}_{\rm SET}\otimes \tilde{\rho}^{1} ,
\label{eq:GIP}
\end{equation}
with $\tilde{\rho}^{0} \equiv \rho^{0}$ and $\tilde{\rho}^{1} \equiv 
\mathrm{diag}\ro{\rho^{1}_{11}, \rho^{1}_{22}}$.  
The Hamiltonian in this interaction picture (after the rotating wave 
approximation) is 
\begin{equation}
\widetilde{H}_{\rm I} = 
  \proj{0}{0}_{\rm SET}\otimes \hat{H}_{\rm qb} 
+ \proj{1}{1}_{\rm SET}\otimes \frac{\varepsilon}{2}\sig_{z} .
\label{eq:HIP}
\end{equation}
Thus we see that in the interaction picture (which reflects the 
\textit{observable} effects of the actual dynamics), no qubit coherence 
exists when the SET is occupied.  That is, the off-diagonal elements of 
${\rho}^{1}$ are equal to zero.

Using the observable state (\ref{eq:GIP}) we obtain the following 
analytical expression for the \textit{observable} coherence function 
\begin{equation}
\label{eq:g_1obs}
g_{\rm obs}(\tau) = P^{0}_{\rm ss} \exp \ro{-\bar{\Gamma}_{\rm L}\tau} ,
\end{equation}
which gives the qubit decoherence time of
\begin{eqnarray}
\tau_{\rm d} &\equiv& \int_{0}^{\infty} g_{\rm obs}(\tau) d\tau
\nn\\
&=& \frac{P^{0}_{\rm ss}}{\bar{\Gamma}_{\rm L}} 
= \frac{\bar{\Gamma}_{\rm R}}
{\bar{\Gamma}_{\rm L}\ro{\bar{\Gamma}_{\rm L}+\bar{\Gamma}_{\rm R}}} .
\label{eq:decotime}
\end{eqnarray}
The result (\ref{eq:g_1obs}) is plotted in \Fig{fig:g} as a bold line.
This indicates that our requirement for 
$\chi\gg\bar{\Gamma}_{\rm L}+\bar{\Gamma}_{\rm R}$ results in an effectively 
immediate loss of a portion 
[$\bar{\Gamma}_{\rm R}/(\bar{\Gamma}_{\rm L}+\bar{\Gamma}_{\rm R})$] 
of the qubit coherence upon commencing measurement, 
followed by exponential decay at the rate $\bar{\Gamma}_{\rm L}$.  

It is interesting to note that in the unlikely event of the SET island state 
being known to be occupied at the start of the measurement (so that $P^{0}=0$), 
the qubit is observed to completely decohere instantaneously.  
This fact alone shows that the quality of charge qubit measurements by a 
SET cannot be captured by a simple quantity such as $\Gamma_{\rm d}\tau_{\rm m}$, 
although this can provide some insight into the measurement quality.  
Note also that $g_{\rm obs}(\tau)$ depends minimally on the SET response.

\subsection{Conditional master equation\label{sec:conditional}}
We obtain the conditional (or selective, or stochastic) evolution of $G(t)$, 
denoted by a subscript c, in a similar way as in Ref. 
\onlinecite{WisetalPRB01}. 
However, we do not eliminate the SET island state as in \Ref{WisetalPRB01}, 
which was achieved by setting 
$\Gamma'_{\rm R}, \Gamma_{\rm R}\gg \Gamma'_{\rm L}, \Gamma_{\rm L}$. 
Our result is the following \ito\cite{Hbook} conditional 
master equation (quantum trajectory/filtering equation), where the SET-qubit 
state is conditioned on the tunneling events both on \textit{and} off the SET 
island:
\begin{eqnarray}
dG_{\rm c}&=&
\sum_{j=\rm L,R} dN_{j}
\cu{\frac{\s_{j}}{\tr{\s_{j}G_{\rm c}}} - 1}G_{\rm c}
\nn\\
&&
-dt~\!\h{\half \hat{\Upsilon} + i\hat{H}}G_{\rm c}\ ,
\label{eq:SME}
\end{eqnarray}
where the non-linear superoperator $\mathcal{H}$ is defined 
by $\h{\hat{x}}G \equiv \hat{x}G + G\hat{x}\dg 
- \tr{\ro{\hat{x}+\hat{x}\dg}G}G$.  
The operator $\hat{\Upsilon}$ and superoperators
$\s_{\rm L}$ and $\s_{\rm R}$ are given by
\begin{subequations}
\begin{eqnarray}
\label{eq:B}
\hat{\Upsilon}   & \equiv &
\Gamma'_{\rm L}\hat{b}\hat{b}\dg \otimes \hat{n}
+ \Gamma_{\rm L}\hat{b}\hat{b}\dg\otimes (\hat{1}-\hat{n})
\nn\\
&& 
~+~ \Gamma'_{\rm R}\hat{b}\dg\hat{b}\otimes \hat{n}
+ \Gamma_{\rm R}\hat{b}\dg\hat{b}\otimes (\hat{1}-\hat{n}) \ , 
\\
\label{eq:SL} \s_{\rm L} & \equiv &
\Gamma'_{\rm L} \J{\hat{b}\dg\otimes\hat{n}}
+ \Gamma_{\rm L}\J{\hat{b}\dg\otimes (\hat{1}-\hat{n})}\ ,
\\
\label{eq:SR} \s_{\rm R} & \equiv &
\Gamma'_{\rm R} \J{\hat{b}\otimes \hat{n}}
+ \Gamma_{\rm R}\J{\hat{b}\otimes (\hat{1}-\hat{n})}\ .
\end{eqnarray}
\label{eqs:convenient}
\end{subequations}

In \Eq{eq:SME} we have introduced the classical random point processes 
$dN_{\rm L}(t)$ and $dN_{\rm R}(t)$.  They represent the number (either 
0 or 1) of electron tunneling events from source to island (L) and island 
to drain (R), respectively, in an infinitesimal time interval $[t,t+dt)$.  
In the sequential tunneling mode, at most one of these may be nonzero for 
each particular infinitesimal interval of time.  That is, an electron cannot 
tunnel through the entire SET `in one go', so to speak.  The conditional 
master equation for the observable state (\ref{eq:GIP}) is identical to 
(\ref{eq:SME}), but with $\widetilde{G}$ in place of $G$ and 
$\widetilde{H}_{\rm I}$ in place of $\hat{H}$.

The two terms summed in the first line of \Eq{eq:SME} represent electron 
tunneling events onto (L) and off (R) the SET island.  Each event is 
represented by an incoherent mixture of the two possible paths by which 
this can occur (corresponding to the two localized electron states within 
the qubit).  The final line in \Eq{eq:SME} represents the evolution of 
$G_{\rm c}(t)$ that is due to the Hamiltonian and the null measurement of 
no SET tunneling events in the time interval $[t,t+dt)$.  It is important 
to note that these microscopic processes may not necessarily be directly 
observed by a realistic observer.\cite{OxtSunWisJPCM03,OxtetalPRB05}  
Nevertheless, it is sensible to unravel the master equation (\ref{eq:ME}) 
into a quantum jump\cite{WisMilPRA93b} stochastic master equation because 
the underlying physical process consists of single electron tunneling 
events.

The ensemble-average evolution of $G(t)$ [\Eq{eq:ME}] can be recovered 
by averaging over all possible trajectories given by \Eq{eq:SME}.  In 
the \ito formalism,\cite{Hbook} this simply involves replacing the 
stochastic increments $dN_{\rm L}$ and $dN_{\rm R}$ in \Eq{eq:SME} 
with their expectation values, which can self-consistently be given in 
terms of their respective quantum averages: 
\begin{equation}
\label{eq:E[dN]}
\mathrm{E}\sq{dN_{\rm L,R}(t)} = {dt~\!} \tr{\s_{\rm L,R}G_{\rm c}(t)} .
\end{equation}

Having presented the quantum trajectory equation (\ref{eq:SME}), 
we are now in a position to consider the SET measurement sensitivity 
using $\tau_{\rm m}$ and $V(t)$ defined in \Sec{sec:meas}.  
We proceed by noting that the qubit state matrix $\rho$ can be 
represented in terms of the Pauli spin matrices as 
\begin{equation}
\rho^{\aleph} = \half\ro{P^{\aleph}\hat{1}
 + x^{\aleph}\sig_{x} 
 + y^{\aleph}\sig_{y}
 + z^{\aleph}\sig_{z}}\ .
\label{eq:bloch}
\end{equation}
The conditional master equation [for the observable state (\ref{eq:GIP})] 
can be expressed in terms of the \textit{conditional} Bloch sphere 
variables $x^{\aleph}_{\rm c}$, $y^{\aleph}_{\rm c}$, and 
$z^{\aleph}_{\rm c}$, as well as the conditional SET island occupation 
probabilities $P^{\aleph}_{\rm c}$.   The conditioning here is upon both 
the measurement results (subscript c) and the SET island state 
(superscript 0 or 1).  To find $V(t)$ (and hence approximate 
$\tau_{\rm m}$), we are interested in the expectation value of 
$(z_{\rm c})^{2} = (z^{0}_{\rm c}+z^{1}_{\rm c})^{2}$.  Averaging over the 
jumps in the conditional master equation for $z_{\rm c}$ and using the 
fact that stochastic variables satisfy $d(z^2) = 2zdz + dzdz$, we obtain 
the following equation for the time rate of change of this expectation 
value:
\begin{eqnarray}
\frac{d\E{z_{\rm c}^{2}}}{dt} &=& 
\mathrm{E}\left[
\frac
{\ro{2\bar{\Gamma}_{\rm L}z_{\rm c}^{0}
 + P^{0}_{\rm c}\Delta \Gamma_{\rm L}}^{2}}
{4P^{0}_{\rm c}\bar{\Gamma}_{\rm L}
 + 2\Delta \Gamma_{\rm L}z_{\rm c}^{0}} 
\right.\nn\\
&& \left.
\phantom{ee}+ \frac
{\ro{2\bar{\Gamma}_{\rm R}z_{\rm c}^{1}
 + P^{1}_{\rm c}\Delta \Gamma_{\rm R}}^{2}}
{4P^{1}_{\rm c}\bar{\Gamma}_{\rm R}
 + 2\Delta \Gamma_{\rm R}z_{\rm c}^{1}}
\right] .
\label{eq:dEz^2}
\end{eqnarray}
Knowing this and using \Eq{eq:qubitsteadystate} to give 
$z^{\aleph}_{\rm c}(0)=0$, we find from \Eq{eq:dotV0} that
\begin{eqnarray}
\tau_{\rm m}^{-1} &\approx &
P^{0}_{\rm ss}
\frac{\ro{\Delta \Gamma_{\rm L}}^{2}}{4\bar{\Gamma}_{\rm L}}
+ P^{1}_{\rm ss}
\frac{\ro{\Delta \Gamma_{\rm R}}^{2}}{4\bar{\Gamma}_{\rm R}}
\nn \\
&=&
\frac{\bar{\Gamma}_{\rm L}\bar{\Gamma}_{\rm R}}
{\bar{\Gamma}_{\rm L}+\bar{\Gamma}_{\rm R}}
\sq{\ro{\frac{\Delta \Gamma_{\rm L}}{2\bar{\Gamma}_{\rm L}}}^{2} 
  + \ro{\frac{\Delta \Gamma_{\rm R}}{2\bar{\Gamma}_{\rm R}}}^{2}} .
\label{eq:locrate}
\end{eqnarray}

\begin{figure}[ht]
\includegraphics[width=0.45\textwidth]{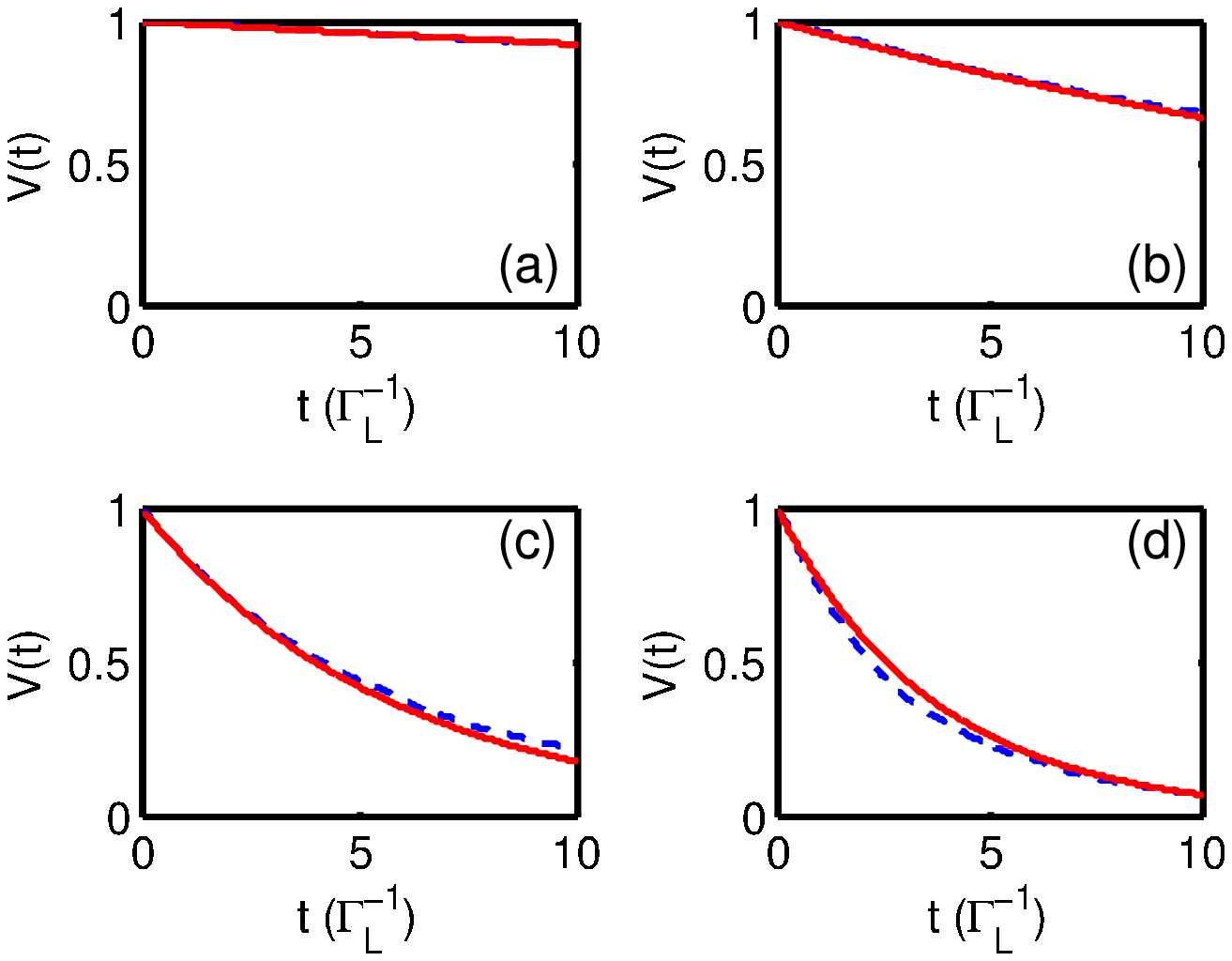}
\caption{The variance in $\sig_{z}$ (the population difference operator 
in the measurement basis), $V(t) = 1 - \E{z_{\rm c}^2(t)}$.  The dashed 
line is an average over 2000 trajectories obtained using the conditional 
master equation.  The solid line is the analytical approximation assuming 
exponential decay of $V(t)$.  Other details are as in \Fig{fig:g}. 
\label{fig:V}}
\end{figure}

A more precise value for the measurement time is obtained from Eq. 
(\ref{eqs:V}) by numerical solution of the conditional master equation 
for the observable state.  
Figure \ref{fig:V} shows plots of $V(t) = 1 - \E{z_{\rm c}^2(t)}$ for a 
strongly coupled SET for the same parameter values 
as the corresponding plots in \Fig{fig:g}.   Notice that the analytical 
approximation (the solid line) agrees very well with all of the 
numerical results (dashed lines).  A comparison of Figs. \ref{fig:g} 
and \ref{fig:V} shows that the strongly responding SET [in plots (d)] 
has comparable decoherence and measurement times because the functions 
$V(t)$ and $g_{\rm obs}(\tau)$ are comparable in magnitude at all times.

\subsection{Quality Of The Measurement \label{sec:quality}}
Previously\cite{SSPRB98,MakSchShnPRL00,KorPRB01b,MakSchShnRMP01,MakSchShn02} 
the quality of the measurement of a charge qubit by a (weakly responding) 
SET has been quantified by comparing qubit decoherence and measurement 
times, with the quantum limit for a weakly responding device given by 
\Eq{eq:hup1}.  

More generally, the purity of the \textit{conditioned} qubit state 
(conditioned by measurement results) provides the means to determine 
the `ideality' of the detector (how close to the quantum limit the 
detector is operating).  This is because an ``ideal'' detector will 
preserve a pure qubit state throughout the measurement.  Thus, we 
maintain that the purity of the conditioned qubit state is the ultimate 
quantifier of measurement quality.  The conditioned state of the qubit 
can only be obtained from the conditional master equation.

In this subsection we analyze the quality of the measurement in two ways.  
We first use the traditional technique of comparing the qubit decoherence 
(back-action) and measurement (sensitivity) times using our definitions 
for these quantities from \Sec{sec:quantify}.  We then analyze the 
average purity of the conditioned qubit state during the measurement.

\subsubsection{Comparison: Measurement time and Decoherence Time
\label{sec:compare}}
Comparing qubit measurement and decoherence time-scales gives an intuitive 
idea of the quality of the measurement.  If these time-scales are of the 
same order, then we can say that the detector is operating close to the 
quantum limit.  Using Eqs. (\ref{eq:decotime}) and (\ref{eq:locrate}) 
we can obtain an analytical approximation for $\Gamma_{\rm d}\tau_{\rm m}$.  
The result is 
\begin{equation}
\Gamma_{\rm d}\tau_{\rm m} \approx 
\frac{\ro{1 + \bar{\Gamma}_{\rm L}/\bar{\Gamma}_{\rm R}}^2}
{\ro{\Delta \Gamma_{\rm L}/2\bar{\Gamma}_{\rm L}}^2
+\ro{\Delta \Gamma_{\rm R}/2\bar{\Gamma}_{\rm R}}^2} .
\label{eq:quality}
\end{equation}
This result is a valid approximation when $V(t)$ is a decaying 
exponential, which is valid for a wide range of parameter values as 
shown in \Fig{fig:V}.  
The minimum for \Eq{eq:quality} is $1/2$, which occurs only for 
$\bar{\Gamma}_{\rm L}\ll\bar{\Gamma}_{\rm R}$ (high SET asymmetry) and 
$\Delta\Gamma_{\rm L,R} \approx 2\bar{\Gamma}_{\rm L,R}$.%
The second condition requires both $\Gamma'_{\rm L}\ll\Gamma_{\rm L}$ 
and $\Gamma_{\rm R}\ll\Gamma'_{\rm R}$, which could in theory be realized 
by careful arrangement of the SET parameters (specifically bias and gate 
voltages, and tunnel junction conductances).

\subsubsection{Average Purity of the Conditioned Qubit State \label{sec:purity}}
\begin{figure}[!ht]
\includegraphics[width=0.45\textwidth]{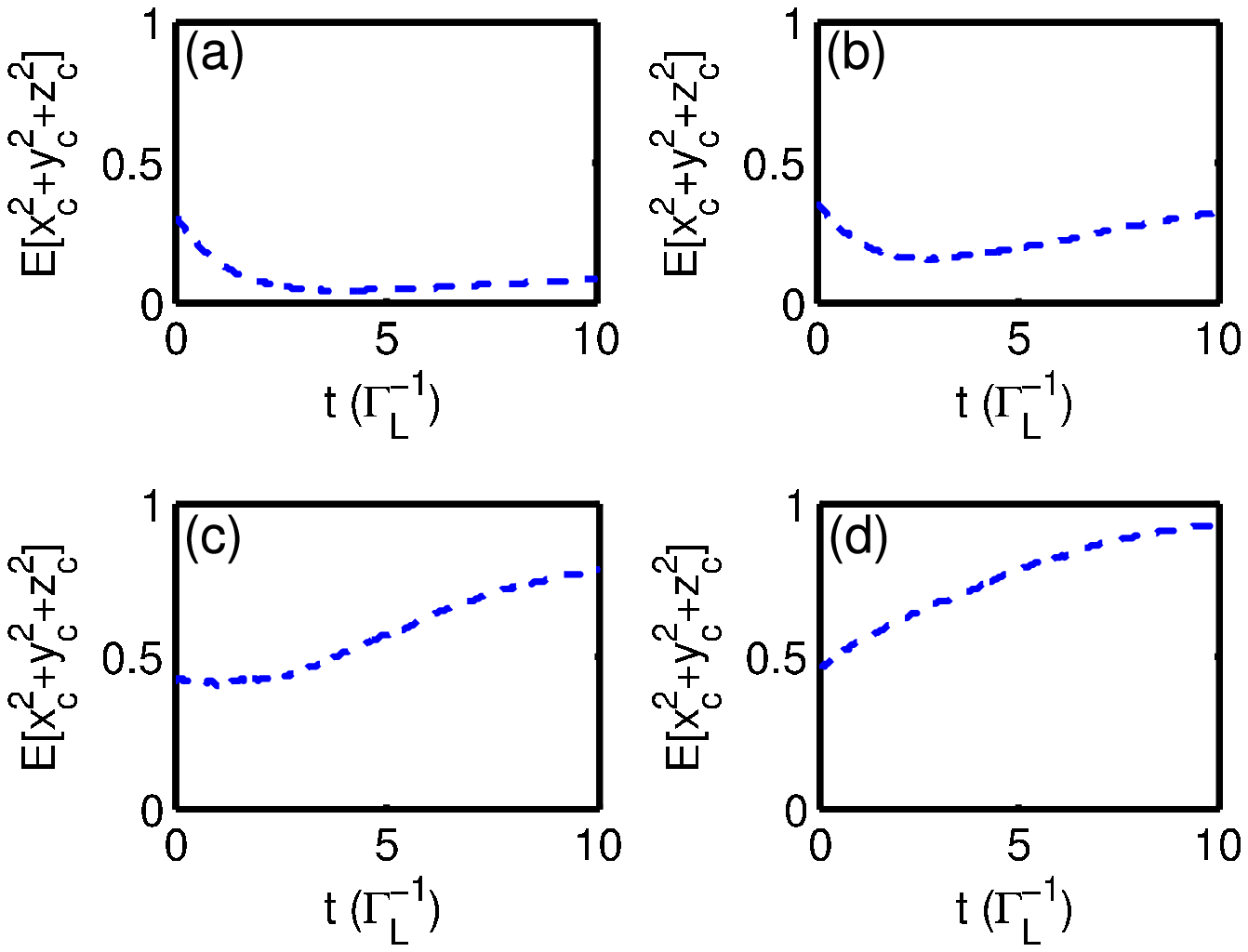}
\caption{Average purity $\E{x_{\rm c}^2+y_{\rm c}^2+z_{\rm c}^2}$ over 2000 
trajectories.  The four plots correspond to the same ensembles calculated 
in \Fig{fig:V}.
\label{fig:purity}}
\end{figure}

Averaging the purity of the conditioned qubit state for the four ensembles 
in \Fig{fig:V} gives the results in \Fig{fig:purity}.  
Here we choose the qubit to start in a pure state [a 
superposition with $z(0)=0$], rather than a mixture.  
At the commencement of the measurement, the purity drops instantaneously 
[as did $g_{\rm obs}(\tau)$].  For weaker response [\Fig{fig:purity} (a) 
and (b)], the purity continues to drop for some time because the SET takes much 
longer to determine the qubit state than it does to collapse it.  This becomes 
apparent by comparing plots (a) and (b) of Figs. \ref{fig:V} and \ref{fig:g}.  
For stronger response the purity approaches 1 rapidly [see \Fig{fig:purity} 
(c) and (d)] because the measurement time is significantly decreased, as can 
be seen in \Fig{fig:V} (c) and (d).  In all four cases shown in 
\Fig{fig:purity}, the continuous measurement eventually purifies the 
conditioned quantum state.\cite{DohJacJunPRA01,FucJacPRA01,KorPB00}

\begin{figure}[ht]
\includegraphics[width=0.45\textwidth]{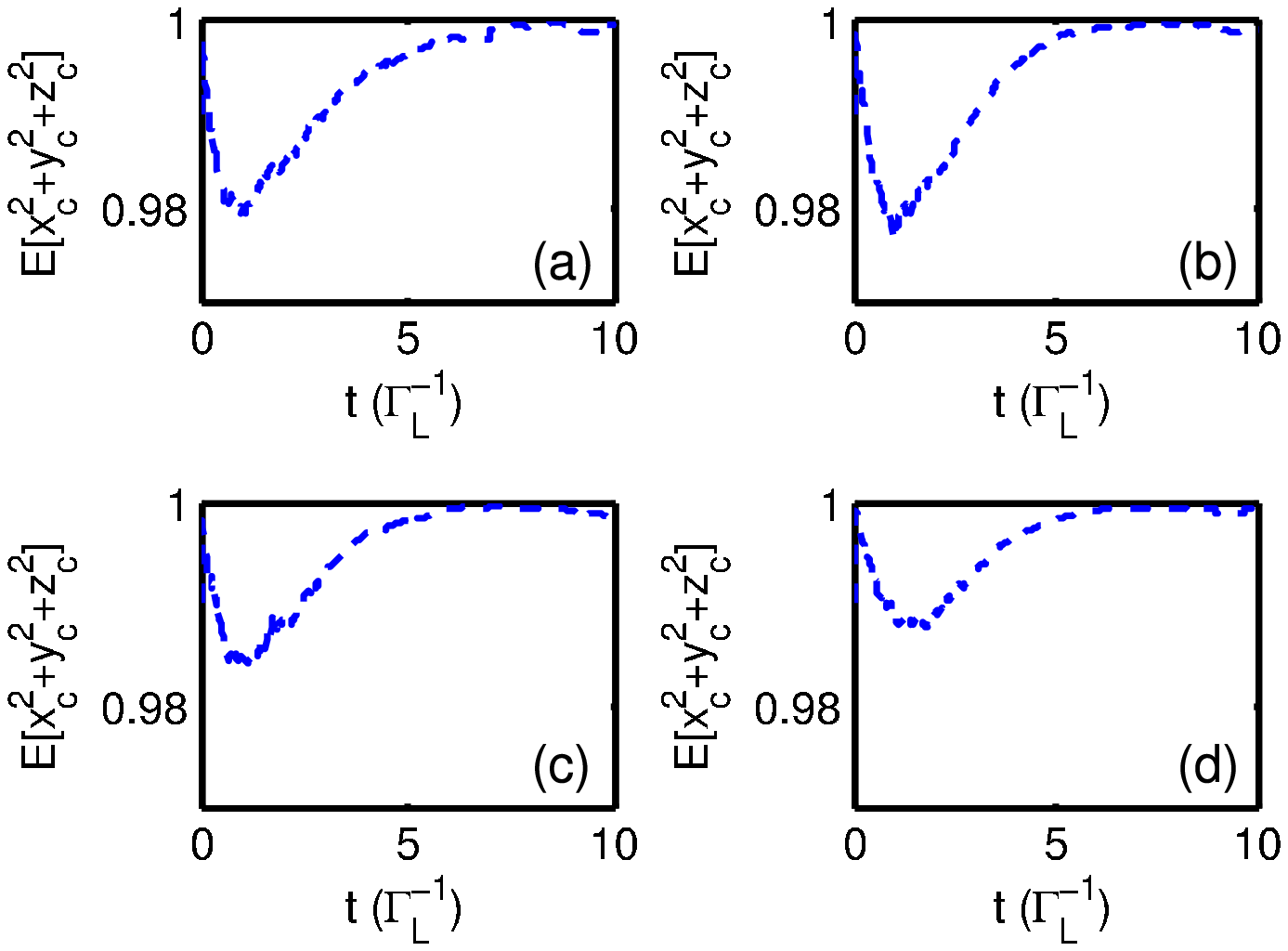}
\caption{Average purity of the conditioned qubit state, 
$\E{x_{\rm c}^2+y_{\rm c}^2+z_{\rm c}^2}$, when monitored by an 
asymmetric SET 
($\bar{\Gamma}_{\rm L} \ll \bar{\Gamma}_{\rm R}$).  
The average is over 2000 trajectories.  Parameter values are given 
in the text.  From (a) to (d), we have 
$\Gamma_{\rm d}\tau_{\rm m} = 1.03$, $0.95$, $0.72$, and $0.59$, 
respectively.
\label{fig:idealpurity}}
\end{figure}

The competing effects of sensitivity and back-action in the qubit measurement 
can be captured by the purity of the conditioned qubit state in the following 
way.  If a pure qubit state remains pure during the measurement, the detector 
is operating at the quantum limit and can be called an ideal\cite{KorPRB99} 
detector.  In order for the qubit to remain 
in a pure state once the measurement starts, 
the SET island must be unoccupied when in steady-state.  
This occurs when electrons tunnel off the island almost immediately, i.e. when 
$\Gamma_{\rm L},\Gamma'_{\rm L} \ll \Gamma_{\rm R},\Gamma'_{\rm R}$.  
This ``asymmetry'' adiabatically eliminates the influence of the SET island 
state on the qubit, as in the model of \Ref{WisetalPRB01}.  
With this condition satisfied, we obtained another ensemble of 2000 quantum 
trajectories using the conditional master equation and plotted the average 
purity of the conditioned qubit state in \Fig{fig:idealpurity}.  
Note the different y-axis limits.  
Again the SET response was controlled by changing $\Gamma'_{\rm L}$ 
[again it was decreased incrementally from (a) to (d)].  The 
SET response was, from (a) to (d), $\Delta I / \bar{I} = 0.23$, $0.58$, 
$1.32$, and $1.70$.  Other parameters for \Fig{fig:idealpurity} were 
$\Omega_0 = 0.01\Gamma_{\rm L}$, 
$\Gamma_{\rm R} = 100\Gamma_{\rm L}$, 
$\Gamma'_{\rm R} = 100\Gamma_{\rm R}$, $\chi = 50\Gamma'_{\rm R}$.  
Similar results were obtained for $\Gamma'_{\rm R} = 10\Gamma_{\rm R}$.  
For the strong response plot of \Fig{fig:idealpurity} (d), this set of 
parameters corresponds to the r\'egime given after \Eq{eq:quality}.  
That is, the r\'egime where the quantity $\Gamma_{\rm d}\tau_{\rm m}$ 
approaches 1/2.  Indeed, the values of $\Gamma_{\rm d}\tau_{\rm m}$ for 
\Fig{fig:idealpurity} approach this limit of 1/2, as given in the figure 
caption.

Figure \ref{fig:idealpurity} shows that the average purity of the 
conditioned qubit state remains very close to 1 throughout the 
measurement for the asymmetric SET.  
Thus we can say that the strongly coupled asymmetric SET operates 
very close to the quantum limit.  This conclusion could not be 
drawn from the values of $\Gamma_{\rm d}\tau_{\rm m}$ alone.  
Interestingly, the SET response [increasing from plot (a) to (d), 
as before] has very little effect on a SET with high asymmetry.  
This is similar to the case of a QPC, which is an ideal detector 
independent of the strength of its response.  
To summarize, the conditions required for a 
strongly coupled SET to operate close to the quantum limit are: 
(a) $0 < \Gamma'_{\rm L} \ll \Gamma_{\rm L}$, and 
(b) $\bar{\Gamma}_{\rm L} \ll \bar{\Gamma}_{\rm R}$.

\section{Conclusion\label{sec:conclusion}}
In this paper we have examined sensitivity and back-action in charge 
qubit measurements by the single-electron transistor (SET) operating 
in the sequential tunneling mode.  
We have shown that the unconditional master equations derived in previous 
work\cite{SSPRB98,MakSchShnPRL00,MakSchShnRMP01,MakSchShn02,GoaPRB04} 
for this measurement (which are equivalent to each other) are not of 
the Lindblad form\cite{LinCMP76} that is necessary and sufficient for 
guaranteeing valid physical states.  Thus, these master equations 
cannot be used to model real physical systems.  In particular, this 
statement also applies to the equation for the conditioned state 
(one that an experimenter would calculate knowing the output of the 
detector) in \Ref{GoaPRB04}.%
\footnote{Note also that Eqs. (21) to (26) of \Ref{GoaPRB04} are 
  incorrectly presented, given the preceding analysis in that reference.  
  Indeed, averaging over the jumps in the conditional master equation 
  [Eqs. (21) and (22) of \Ref{GoaPRB04}] results in an unconditional 
  master equation different from that presented [the average of Eq. (10) 
  of \Ref{GoaPRB04} over the drain of the SET].  We found no principle 
  errors in the rest of the unraveling process presented prior to the 
  conditional master equation.
  }  
The unconditional master equation was derived independent of the 
strength of SET-qubit coupling, but analysis of the SET measurement 
quality (or efficiency) 
assumed\cite{SSPRB98,MakSchShnPRL00,MakSchShnRMP01,MakSchShn02} 
weak SET response.  Weak response is when the average component of the 
SET output is much larger than the component that depends on the 
qubit state.

We have presented an alternative master equation that \textit{is} of 
the Lindblad form, as well as the associated quantum trajectory 
equation\cite{OpenSys,WisMilPRA93b,WisQSO96} for the conditioned 
state.  To derive these equations required assuming strong SET-qubit 
coupling.  A strongly coupled detector can exhibit strong response 
or weak response, whereas weak coupling implies weak response.

We have shown that, in general, quantifying the quality of charge qubit 
measurements by a SET is not as simple as merely comparing the qubit 
decoherence (also called coherence) time and measurement time (as done 
previously for the weakly responding SET in Refs. \onlinecite{SSPRB98}, 
\onlinecite{MakSchShnPRL00}, \onlinecite{MakSchShnRMP01}, and 
\onlinecite{MakSchShn02}).  This is because, for strong SET-qubit 
coupling, the SET island charge state has a large effect on the qubit 
coherence.  Despite this, comparing qubit decoherence and measurement 
times does provide some insight into the measurement quality.

Previous definitions for the qubit decoherence time%
\cite{SSPRB98,MakSchShnPRL00,KorPRB01b,MakSchShnRMP01,MakSchShn02} 
either produced an overestimate of the measurement quality (as in 
Refs. \onlinecite{SSPRB98}, \onlinecite{MakSchShnPRL00}, 
\onlinecite{MakSchShnRMP01}, and \onlinecite{MakSchShn02}), 
or are only valid for a weakly responding detector (as in \Ref{KorPRB01b}).  
We presented a new definition for the decoherence time that is
valid for strong SET-qubit coupling and arbitrary response, by 
adopting a coherence function technique\cite{QNoise} from the 
field of quantum optics.

Previous definitions for the measurement time%
\cite{SSPRB98,MakSchShnPRL00,KorPRB01b,MakSchShnRMP01,MakSchShn02} 
are only valid for weakly responding detectors.  Thus, we also 
presented a new definition of the measurement time which is analogous 
to the new decoherence time definition.  Our definition is in terms of the 
conditioned qubit state, and involves an average over many stochastic 
measurement records.  An analytical approximation for the measurement 
time compared well to the numerical solution.  A similar approach to 
quantifying the measurement sensitivity was first considered in 
\Ref{WisetalPRB01}.

An important result of this paper was to show that the strongly coupled 
SET can operate at, or very near, the quantum limit during charge qubit 
measurements.  This was shown in two ways.  The first involved comparing 
the decoherence and measurement times found using our Lindblad-form 
master equations.  In the strong \textit{response} limit, these times 
were found to be of the same order as each other.  For the highly 
asymmetric SET where tunneling into the drain (collector) occurs at a 
much higher rate than tunneling onto the SET from the source (emitter), 
these times were of the same order as each other, independent of the 
SET response.  This is also the case for the QPC, which can be operated 
as an ideal detector.\cite{KorPRB99}  The second, and our preferred, 
way to show that the strongly coupled SET can operate close to the 
quantum limit was by considering the average qubit purity for an 
ensemble of stochastic measurement records.  Again the asymmetric SET 
displayed near ideal detector behavior by maintaining the purity of the 
qubit state above $98\%$.

As far as we are aware, our Lindblad-form quantum trajectory equation 
is the first of its type presented for charge qubit measurement by a 
SET.  Moreover, having shown that the SET can approach the quantum 
limit for charge qubit measurements, it is now meaningful to study 
how the measurement is affected by extra classical noises and 
filtering due to a realistic measurement 
circuit.\cite{OxtSunWisJPCM03,OxtetalPRB05}  
This research is continuing.

\acknowledgments
This research was supported by the Australian Research Council 
and the State of Queensland.  NPO thanks Joshua Combes, 
Steve Jones and Jay Gambetta for useful discussions.  
HS acknowledges F. Green for valuable discussions.

\appendix 
\section{Sensitivity and Back-action in a QPC Measurement\label{appQPC}}
The qubit decoherence rate for measurement by a low-transparency QPC 
(single tunnel junction device) can be obtained either by inspection of 
the master equation,%
\cite{GurPRB97,KorPRB99,GoaMilWisSunPRB01} or using the technique 
introduced in \Sec{sec:deco}.  The same result is obtained using 
both methods:
\begin{equation}
\label{eq:decorate_qpc}
\Gamma_{\rm d} = \frac{\ro{\sqrt{\Gamma} - \sqrt{\Gamma'}}^{2}}{2} ,
\end{equation}
where $\Gamma$ ($\Gamma'$) is the tunneling rate through the QPC when 
the target dot is unoccupied (occupied).  Using the new technique 
introduced in \Sec{sec:meas}, the analytical approximation of the 
qubit measurement time (which agrees with \Ref{GoaMilWisSunPRB01})
is
\begin{equation}
\tau^{-1}_{\rm m} = \frac{\st{\Delta \Gamma}^2}{4\bar{\Gamma}}
= \frac{\ro{\sqrt{\Gamma} +
\sqrt{\Gamma'}}^{2}}{\Gamma + \Gamma'}\Gamma_{\rm d} .
\label{eq:locrate_qpc}
\end{equation}
Combining these results gives
\begin{equation}
\Gamma_{\rm d}\tau_{\rm m} 
= \label{eq:qpcquality1}
\frac{\Gamma + \Gamma'}{\ro{\sqrt{\Gamma} + \sqrt{\Gamma'}}^{2}} .
\end{equation}
Note that $1 \geq \Gamma_{\rm d}\tau_{\rm m} \geq 1/2$.  The upper 
bound is approached for a ``strongly responding'' QPC 
($\Gamma \gg \Gamma'$).  The lower bound is approached for a ``weakly 
responding'' QPC ($\Gamma \approx \Gamma'$), i.e. when the measurement 
and decoherence rates go to zero.

\section{The SET-qubit Master Equation of Goan\label{sec:goanme}}
In \Ref{GoaPRB04}, Goan presents a master equation for the combined 
state of the qubit plus island and drain of the SET.  In that paper, 
the master equation is shown to be equivalent to that in 
\Ref{MakSchShnPRL00}.  
Using our notation and tracing over the drain degrees of freedom gives
\begin{eqnarray}
\dot{G} & = & 
-i\sq{\hat{H},G}
+ \Gamma'_L\D{\hat{b}\dg\otimes\hat{n}}G
+ \Gamma'_R\D{\hat{b}\otimes\hat{n}}G 
\nn\\ &&
+ \Gamma_L\D{\hat{b}\dg\otimes\ro{1-\hat{n}}}G
+ \Gamma_R\D{\hat{b}\otimes\ro{1-\hat{n}}}G
\nn\\ &&
+ 2\bar{\Gamma}_{\rm L}\D{\hat{n}}\hat{b}\dg G \hat{b}
+ 2\bar{\Gamma}_{\rm R}\D{\hat{n}}\hat{b}G \hat{b}\dg ,
\label{eq:goanme2}
\end{eqnarray}
where $\hat{b}\dg$ corresponds to $e^{i\phi}$ in \Ref{GoaPRB04}, and we 
have used $\sq{\hat{n},\sq{\hat{n},\hat{x}}} = -2\D{\hat{n}}\hat{x}$.  
This master equation is equivalent to those in Refs. 
\onlinecite{SSPRB98}, \onlinecite{MakSchShnPRL00}, 
\onlinecite{MakSchShnRMP01}, and \onlinecite{MakSchShn02}.  
Comparison of this equation with our unconditional master equation 
(\ref{eq:ME}) shows that the terms in the final line of \Eq{eq:goanme2} 
are not present in our model.  These are the terms that lead to the 
non-Lindbladian dynamics in Refs. \onlinecite{SSPRB98}, 
\onlinecite{MakSchShnPRL00}, and 
\onlinecite{MakSchShnRMP01,MakSchShn02,GoaPRB04}.

\bibliography{oxtwissun}

\end{document}